\documentclass[12pt, letterpaper]{JHEP3}
\usepackage{amsmath,epsf,amssymb,latexsym,cite,graphics,bbm}

\DeclareFontFamily{U}{rsf}{}
\DeclareFontShape{U}{rsf}{m}{n}{
  <5> <6> rsfs5 <7> <8> <9> rsfs7 <10-> rsfs10}{}
\DeclareMathAlphabet\Scr{U}{rsf}{m}{n}

 \numberwithin{equation}{section}

\def\al{\alpha}
\def\da{{\dot\alpha}}
\def\db{{\dot\beta}}
\def\bet{\beta}

\def\eps{\epsilon}

\def\tha{\theta}
\def\th{\theta}
\def\sig{\sigma}

\def\r{\rho}

\def\Om{\Omega}
\def\Sig{\Sigma}

\newcommand{\bea}{\begin{eqnarray}}
\newcommand{\eea}{\end{eqnarray}}
\newcommand{\be}{\begin{equation}}
\newcommand{\ee}{\end{equation}}

\newcommand{\cG}{{\cal G}}

\def\cJ{{\cal J}}

\def\cO{{\cal O}}

\def\T{{\cal T}}

\def\cV{{\cal V}}

\newcommand{\RR}{{\mathbb R}}

\newcommand{\half}{\frac{1}{2}}
\newcommand{\qrt}{\frac{1}{4}}
\newcommand{\del}{\partial}

\newcommand{\delbar}{\bar\partial}

\newcommand{\SO}{\operatorname{SO}}

\newcommand{\w}{\wedge}

\newcommand{\ol}{\overline}

\newcommand{\bbar}{{\ol{b}}}

\newcommand{\ibar}{{\ol{\imath}}}
\newcommand{\jbar}{{\ol{\jmath}}}
\newcommand{\kbar}{{\ol{k}}}

\newcommand{\qbar}{{\ol{q}}}

\newcommand{\Qbar}{\ol{Q}}

\newcommand{\Ombar}{\ol{\Om}}

\newcommand{\dal}{{\dot{\al}}}
\newcommand{\thab}{\bar{\theta}}
\newcommand{\thal}{\theta^{\al}}

\newcommand{\non}{\nonumber}

\newcommand{\bpm}{\begin{pmatrix}}
\newcommand{\epm}{\end{pmatrix}}
 \newcommand{\bitem}{\begin{itemize}}
 \newcommand{\eitem}{\end{itemize}}

\newcommand{\C}[1]{$(\ref{#1})$}

\def\Vol{\operatorname{Vol}}

\def\cc{\text{c.c.}}

\def\SO{\operatorname{SO}}

\def\SU{\operatorname{SU}}
\def\U{\operatorname{U}}

\def\e{\epsilon}
\def\bz{\bar{z}}

\def\Vext{{\mathcal{V}}_{d=4}}
\def\Vint{{\mathcal{V}}_{\rm int}}

\title{Type IIB Flux Vacua from the String Worldsheet}
\author {William D. Linch {\sc iii}\\
\normalsize C.\,N.\,Yang Institute for Theoretical Physics and\\ Department of Mathematics,\\
\normalsize State University of New York,\\
\normalsize Stony Brook, NY 11794-3840, USA.\\
Email:  \email{wdlinch3@math.sunysb.edu}}
\author{ Jock McOrist \\
\normalsize Enrico Fermi Institute, \\
\normalsize University of Chicago, \\
\normalsize Chicago, IL 60637, USA.\\
Email: \email{jmcorist@uchicago.edu}}
\author{  Brenno Carlini Vallilo \\
\normalsize Instituto de F\'\i sica,\\ Universidade de S\~ao Paulo,\\
\normalsize C.P. 66318, 05315-970, S\~ao Paulo, SP, Brasil.\\
Email: \email{vallilo@fma.if.usp.br}}

\abstract{We study type IIB string compactifications in the presence of RR and NSNS fluxes using worldsheet techniques. Vertex operators corresponding to internal RR and NSNS fluxes are constructed in the hybrid formalism with manifest target space supersymmetry. In a simple class of examples, we compute various known physical phenomena such as warping and the generation of a superpotential for complex structure moduli. The results are in precise agreement with the supergravity literature.
}

\preprint{EFI-08-08\\
YITP-SB-08-13}
\keywords{Flux compactifications, superstring vacua, hybrid formalism}

\begin{document}


\section{Introduction}
The study of string vacua is an important direction in string theory both from the point of view of constructing realistic phenomenology and in the context of ideas such as the `landscape' of string vacua. To that end, it is important to understand how fluxes, especially Ramond-Ramond (RR) fluxes, affect the existence and modify the properties of string solutions. At low energies and at large volume, where $\alpha'$ corrections are under control, one can study string solutions using the techniques of supergravity. Already from the point of view of supergravity it is clear that RR vacua have some interesting features: Moduli are lifted and the space back-reacts, becoming warped. At string scale, $\alpha'$ corrections will modify the properties and existence of string solutions. For example, there is evidence for the existence of `non-geometric' solutions of string theory that have no large volume limit (see \cite{Wecht:2007wu} and references therein). To obtain a better understanding of such string scale solutions, one needs a string worldsheet description of RR flux vacua.



Studying flux backgrounds on the worldsheet has traditionally proven
difficult. In the Ramond-Neveu-Schwarz (RNS) formulation, RR
vertex operators carry non-zero picture charge and have branch cuts in their operator product expansions,
due to which, one cannot easily exponentiate them into the worldsheet action.
The Green-Schwarz (GS) formalism circumvents some of these problems
although the action cannot be covariantly quantized. Thus the GS approach is
limited to a classical analysis and to backgrounds that admit light cone
gauge. Alternative formalisms developed by Berkovits and
collaborators have the nice feature of manifest space-time supersymmetry and
admit a covariant quantization.
These include the pure spinor
formalism\cite{Berkovits:2000fe} and the hybrid formalism in
$d=2$\cite{Berkovits:2001tg}, $d=4$\cite{Berkovits:1994wr} and
$d=6$\cite{Berkovits:1994vy,Berkovits:1999du} which can be used to
study RR backgrounds, as we do in this paper.

Previous
studies of RR backgrounds have been undertaken in the context of
$AdS_3\times S^3$\cite{Berkovits:1999im}, $AdS_2 \times
S^2$\cite{Berkovits:1999zq}, the C-deformation\cite{Ooguri:2003qp}, and
non-commutative superspace \cite{Seiberg:2003yz,deBoer:2003dn}. In all of
these settings, the RR flux lies entirely in the uncompactified sector. RR
fluxes in the compactified sector have been considered in
\cite{Vafa:2000wi,Berkovits:2003pq,Lawrence:2004zk,Linch:2006ig}. For
example, in \cite{Vafa:2000wi} and later in
\cite{Berkovits:2003pq,Lawrence:2004zk}, the relation between internal
fluxes, auxiliary fields, and the hybrid vertex operators was discussed\footnote{After this work was completed, we were informed that in unpublished work, \cite{Sethi} obtained some related results in the context of type IIA on a Calabi-Yau $4$-fold using the $d=2$ Hybrid formalism of \cite{Berkovits:2001tg}. This included computing RR flux vertex operators, and computation of correlation functions which implied the presence of a superpotential.}. In
\cite{Berkovits:2003pq,Lawrence:2004zk} it was pointed out that the proposal
for these flux vertex operators did not satisfy certain physical state
conditions so it was not clear how to consistently construct flux
deformations of the worldsheet action. In \cite{Linch:2006ig}, a modified
flux vertex operator was proposed that satisfied the physical state
conditions. Open questions remained, however,
concerning the precise physical meaning of the flux vertex
operator and a systematic study of the associated space-time
physics.

In this paper, we clarify and expand the presentation in \cite{Linch:2006ig}. We identify the vertex operators corresponding to both Ramond-Ramond and Neveu-Schwarz-Neveu-Schwarz (NSNS) fluxes in the hybrid formalism and resolve the issues alluded to in the previous paragraph. This is achieved by mapping the RNS flux vertex operators to the hybrid formalism via the auxiliary fields present in the $N=2$ superspace formalism of \cite{Berkovits:1995cb}. This is an easier route than directly applying the RNS-hybrid map discussed in \cite{Berkovits:1996bf,Berkovits:1994vy,Berkovits:1993xq,Berkovits:1993zy}. Using our identification, we compute some of the physics associated with this flux deformation in the context of a simple, non-compact, conformally Calabi-Yau background. We study the effect of non-compactly supported fluxes in perturbation theory. This background also has a well-defined supergravity description with which we can compare our string calculations. Even in perturbation theory, we find many effects known from the supergravity literature. For example, we show how to derive warping (see also \cite{Linch:2006ig}) and a superpotential for complex structure moduli. All of these effects are in precise agreement with the existing supergravity literature. 

The outline for this paper is as follows. In section \ref{sect:cy_1} we give a schematic overview of the hybrid formalism, comparing and contrasting it with the RNS formulation. In section \ref{sect:deformations} we identify the type IIB vertex operators for RR and NSNS flux deformations in the hybrid formalism. In section \ref{sect:applications} we compute some simple physical effects and compare with known supergravity calculations. In section \ref{sect:conclusions} we summarize our results and give an outlook on future work.

\section{The Hybrid Formalism on Calabi-Yau 3-folds}
\label{sect:cy_1}
In preparation for the discussion of flux deformations of Calabi-Yau compactifications of type IIB strings to 4 dimensions, we highlight some basic features of the $d=4$ hybrid string worldsheet and contrast it with that of the RNS formulation (more details can be found in \cite{Berkovits:1994wr,Berkovits:1995cb,Berkovits:1996bf}).

\subsection{Aspects of the String Worldsheet}
The (left-moving part of the) RNS formalism has a field content consisting of worldsheet bosons $X^M$ and Majorana-Weyl fermions $\psi^M$ (where $M=0,1\ldots,9$) together with superconformal ghosts $(b,c,\beta,\gamma)$. The formulation has a manifest $N=(1,1)$ worldsheet supersymmetry. This comes at the expense of manifest space-time supersymmetry: In this formalism, space-time supersymmetry requires the introduction of picture changing and complicated spin-fields with branch cuts in their operator products. This means that one cannot readily study RR backgrounds since the RR vertex operators are constructed from the spin-fields which cannot be easily exponentiated into the action.

The other major formulation of the string worldsheet theory is the Green-Schwarz formalism in which the fundamental field content consists of the worldsheet bosons $X^M$ and Grassmann variables $\thal_L,\thal_R$. In this approach, target-space supersymmetry is realized linearly at the cost of manifest worldsheet supersymmetry. The Green-Schwarz formalism can classically describe fermionic and RR fields but is difficult to quantize due to the presence of second-class constraints inherent in its formulation. In order to quantize the worldsheet theory, one typically breaks Lorentz invariance ({e.g.}~via light-cone gauge). The gauge choice complicates string amplitudes and restricts one to backgrounds that are compatible with the gauge choice. By contrast, important classes of flux backgrounds, including the ones we will consider, contain a space-time warp factor which is typically not consistent with the gauge choice. Thus, it is desirable to have a covariantly quantizable formulation in which space-time supersymmetry is manifest and RR fields can be treated naturally. For string compactifications preserving $\SO(3,1)$ Lorentz symmetry, the 4-dimensional hybrid formalism proposed by Berkovits \cite{Berkovits:1994wr} is a natural candidate.


Before discussing flux deformations, we review the hybrid formalism without flux as originally proposed in \cite{Berkovits:1994wr}. The 10-dimensional space-time manifold is taken to be of the form
\be
\RR^{3,1} \times M_6,\label{eqn:target_space_1}
\ee
where $M_6$ is an internal manifold which preserves $N=2$ space-time supersymmetry. In such a background the worldsheet theory factorizes. In general, the worldsheet theory will not factorize and the target space-time will not be a direct product as in \C{eqn:target_space_1} as discussed in section \ref{sect:applications}.

The hybrid formalism can be understood as a field redefinition of RNS. Take $M_6=CY_3$ to be a Calabi-Yau 3-fold. Then the RNS worldsheet action splits as
\be
S_{RNS} = S_{d=4} + S_{CY_3} + S_{\rm ghosts},\label{eqn:action_1}
\ee
where $S_{d=4}$ describes the superconformal field theory parameterized by $X^\mu,\psi^\mu$, and $S_{CY}$ describes an $N=(2,2)$ superconformal field theory. We will parameterize the internal Calabi-Yau manifold by the complex coordinates $y^i,\bar y^\ibar$ and denote the corresponding complex Weyl fermions by $\psi_{L,R}^i,\bar{\psi}_{L,R}^\ibar$. The hybrid variables are obtained by performing a field redefinition on the space-time variables and ghost content together. That is, one replaces
\be
(x^\mu,\psi^\mu,b,c,\beta,\gamma)\rightarrow (x^\mu, \theta_{L}^\al, \bar \theta_{L}^{\da},p_{L\,\al},\bar p_{L\,\da},\rho_{L}),
\ee
and similarly for the right-moving sector, where $(\theta_{L,R},\bar\theta_{L,R})$ are the usual $4d$, $N=2$ Grassmann superspace coordinates, $(p_{L,R},\bar p_{L,R})$ are their conjugate momenta, and $(\r_L,\r_R)$ are two chiral ghost bosons. The spinor index indicates the space-time chirality. The $(p,\theta)$ variables are a $(b,c)$-type system with weights $1$ and $0$ respectively. The resulting action is a combination of GS in the 4-dimensional space and RNS in the internal space taking the form
\be
S_{\rm hybrid} = S_{GS,d=4} + S_{CY} + S_{\rm chiral}.\label{eqn:action_2}
\ee
Some general comments on this construction are:
\begin{itemize}
\item The space-time part of the target manifold is the standard $4d$, $N=2$ superspace with coordinates, and 4-dimensional super-Poincar\'e symmetry is manifest. This allows a unified description of fermions and RR and NSNS fields.
\item The action is free in a flat background and easily quantizable, lacking the second class constraints which complicate the quantization of the GS string.
\item All the ghost content is now tied up in the definition the GS variables $(p,\theta)$ and the chiral bosons $\rho_{L,R}$ so there is no need to introduce additional ghosts in order to describe scattering processes.
\item The field content is automatically GSO projected, and there are no branch cuts. This eliminates the difficult sum over spin structures essential for target space supersymmetry in RNS.
\item The hybrid formalism has been subjected to many non-trivial tests
and has passed them all. For example, the spectra of RNS and the hybrid
formalisms match, as well as the tree \cite{Berkovits:1996cr} and
1-loop\cite{Berkovits:2001nv} amplitudes. It was shown that superconformal
invariance at quantum level implies the correct four dimensional supergravity
equations of motion for both heterotic and type IIB case
\cite{skend,nedel}.
\end{itemize}
We will now discuss the space-time sector (and chiral bosons) and then turn to a discussion of the internal (Calabi-Yau) sector.

\subsection{Four-dimensional Sector}
\noindent The 4-dimensional space-time superconformal field theory is described by a free field action for the GS like variables. The action is
\bea
S_{GS,d=4} &=& \int d^2z \left[\half \del_L X^\mu \del_R X_\mu + p_{L\al}\del_R \theta_L^{\al} + \bar p_{L\da}\del_R \bar \theta_L^{\da} \right].
\eea
The the resulting operator products are
\bea
X(z)^\mu X(0)^\nu &\sim& -\eta^{\mu\nu} \ln |z|^2,\non\\
p_{L\al} (z)\theta_L^\beta (0)\sim \frac{\delta^\beta_\al}{z}, &\quad& \bar p_{L\da}(z) \bar \theta^{\db}_L(0) \sim \frac{\delta_\da^\db}{z},\non\\
p_{R\al} (\bz)\theta_R^\beta (0)\sim \frac{\delta^\beta_\al}{\bz}, &\quad& \bar p_{R\da}(\bz) \bar \theta^{\db}_R(0) \sim \frac{\delta_\da^\db}{\bz},
\label{eqn:algebra_1}
\eea
while the chiral bosons have the operator products
\bea
\rho_L(z)\rho_L(0) \sim -\ln z, &\quad& \rho_R(\bz)\rho_R(0)\sim - \ln \bz .\label{eqn:algebra_1b}
\eea
The space-time supersymmetry charges are given by
\bea
Q_{L\al}=\oint dz\left(p_{L\al}-\frac{i}{2}(\sig^m\thab_L)_{\al}\del_L
X_m-\frac{1}{8}(\thab_L)^2\del_L\tha_{L\al}\right),\cr
\Qbar_{L\dal}=\oint
dz\left(\bar{p}_{L\dal}-\frac{i}{2}(\tha_L\sig^m)_{\dal}\del_L
X_m-\frac{1}{8}(\tha_L)^2\del_L\thab_{L\dal}\right),\cr
Q_{R\al}=-\oint
d\bar{z}\left(p_{R\al}-\frac{i}{2}(\sig^m\thab_R)_{\al}\del_R
X_m-\frac{1}{8}(\thab_R)^2\del_R\tha_{R\al}\right),\cr
\Qbar_{R\dal}=-\oint
d\bar{z}\left(\bar{p}_{R\dal}-\frac{i}{2}(\tha_R\sig^m)_{\dal}\del_R
X_m-\frac{1}{8}(\tha_R)^2\del_R\thab_{R\dal}\right).\label{eqn:algebra_2}
\eea
The set of Green-Schwarz-Siegel operators which commute with these charges are
\cite{Siegel:1985xj}
\bea
d_{L\al} &=& p_{L\al} + \frac{i}{2}(\sig^m\thab_L)_{\al}\del_L
X_m - \frac{1}{4} (\bar\theta_L)^2 \del_L\theta_{L\al} + \frac{1}{8} \theta_{L\alpha} \del_L(\bar\theta_L)^2,\non\\
\bar d_{L\da} &=& \bar p_{L\da} + \frac{i}{2}(\tha_L\sig^m)_{\dal}\del_L
X_m - \frac{1}{4} (\theta_L)^2 \del_L\bar\theta_{L\da} + \frac{1}{8} \bar\theta_{L\da} \del_L(\theta_L)^2,\non\\
\Pi_L^\mu &=& \del_L X^\mu - \frac{i}{2} \sigma^\mu_{\al\da} \left(\theta_L^\al \del_L \bar\theta_L^\da + \bar\theta_L^\da \del_L\theta_L^\al \right),\label{eqn:algebra_3}
\eea
together with their right-moving counterparts. The operator product expansion (\ref{eqn:algebra_1}) implies that $d_L$ and $\bar d_L$ act on vertex operators as superspace covariant derivatives (c.f. equation (\ref{eqn:CovDer})) and that these operators satisfy the algebra
\bea
d_{L\al}(z) \bar d_{L\da}(0) &\sim& i \frac{\Pi_{L\al\da}}{z}, \quad d_{L\al}(z) d_{L\beta}(0) \sim {\rm regular}, \quad \bar d_{L\da}(z) \bar d_{L\da}(0) \sim {\rm regular},\non\\
d_{L\al}(z) \del_L \theta_L^\beta(0) &\sim& \frac{\delta_\alpha^\beta}{z^2}, \quad \bar d_{L\da}(z) \del_L \bar\theta_L^{\db}(0) \sim \frac{\delta_\da^\db}{z^2},\non\\
d_{L\al}(z)\Pi^\mu_L(0) &\sim& -i \frac{\sigma^\mu_{\al\da} \del_L \bar\theta_L^\da(0)}{z}, \quad \bar d_{L\da}(z) \Pi_L^\mu(0) \sim -i \frac{\sigma^\mu_{\al\da}\del_L\theta_L^\al(0)}{z},\non\\
\Pi_L^\mu(z) \Pi_L^\nu(0) &\sim& - \frac{\eta^{\mu\nu}}{z^2}.\label{eqn:algebra_4}
\eea
There are analogous operator product expansions for the right movers.

The 4-dimensional space-time sector actually has a non-linear $N=(2,2)$ superconformal symmetry with central charge $\hat c=-1$. The operators that generate it are given by
\bea
T_{L, d=4} &=& -\frac{1}{2} \del_L X^\mu \del_L X_\mu - p_{L\al}\del_L \theta^\al_L - \bar p_{L\da} \del_L\bar\theta_L^\da - \half \del_L\rho_L \del_L \rho_L,\non\\
G_{L, d=4}^+ &=& e^{\rho_L}(\bar d_L)^2, \quad G_{L, d=4 }^- = e^{-\rho_L}(d_L)^2, \quad J_{L, d=4} = -\del_L \rho_L,\label{eqn:algebra_5}
\eea
again, with analogous definitions for the right-moving side. The origin of this $N=(2,2)$ supersymmetry lies in the relation between the parent RNS theory and hybrid reformulation: By requiring BRST and conformal invariance of the RNS string (with a gauged  $N=(1,1)$ supersymmmetry) one can show that the RNS string possesses a twisted $N=(2,2)$ supersymmetry \cite{Berkovits:1993xq}. This symmetry is preserved by the hybrid field redefinition and takes the form \C{eqn:algebra_5}. We will use this $N=(2,2)$ supersymmetry as a guiding principle when we discuss allowable flux deformations.

In the hybrid variables only $8$ space-time supersymmetries are realized linearly.  If the background preserves a larger number of supersymmetries ({e.g.}~if $M_6 = T^6$) the remaining supersymmetries will be non-linearly realized. We will return to this when we discuss the $F_5$ flux in section \ref{sect:f1_f5}.

We are now free to couple this $\hat c = -1$ superconformal field theory describing the space-time part of the target-space to a $\hat c= 3$ internal superconformal field theory. This gives a critical $\hat c =2$, $N=2$ string. To compute scattering amplitudes one could, in principle, add $\hat c =-2$ superconformal ghosts and perform BRST quantization. This was originally done in \cite{Berkovits:1993zy} and the results were shown in \cite{Berkovits:1993xq} to coincide with the $N=1$ RNS string. However, it is easier to use the techniques described in \cite{Berkovits:1994vy} to embed this theory into an $N=4$ topological theory after which one can compute amplitudes using the rules for topological strings. This is our approach in section \ref{sect:superpontetial}.

\subsection{Internal Calabi-Yau Sector}
The internal sector is described by a $\hat c =3$, $N=(2,2)$ superconformal field theory. The generators of the superconformal algebra will be denoted by $(\T,\cG_L^+,\cG_L^-,\cJ_L)$ where $\cJ_L$ is the $\U(1)$ $R$-symmetry generator.
In section \ref{sect:deformations} the correspondence between chiral ring elements and harmonic forms will be used to identify worldsheet operators corresponding to fluxes on $M_6$.
In order to present a coherent picture, this identification will now be briefly reviewed for a Calabi-Yau target space. (Some nice reviews of $N=(2,2)$ theories and chiral rings include \cite{Warner:1993zh,Greene:1996cy}.)

An operator $\Phi$ is said to be a {\it chiral primary} operator if it is primary operator ($\mathcal{T}_n \cdot \Phi = \cG^\pm_r\cdot\Phi = 0$ for $n,r>0$) and, in addition, satisfies
\be
(\cG^+_{L})_{-\half} \cdot \Phi = 0.
\ee
Similarly, an anti-chiral operator satisfies the same conditions with $(\cG^-_L)_{-\half}$ replacing $(\cG^+_L)_{-\half}$. The set of chiral primary fields forms a ring under operator products, and chiral primary operators in a unitary representation saturate the bound $h\ge \half q$ (where $h$ is the conformal weight and $q$ the $\U(1)_L$ charge). Analogous statements apply to the right-moving sector. The set of primaries that are chiral with respect to both the left- and right-moving algebra form the $(c,c)$-ring while elements of the $(c,a)$-ring are twisted-chiral primaries. Up to complex conjugation, these are the two possible chiral rings for a given $N=(2,2)$ theory. In addition to chiral and twisted-chiral multiplets, there are also semi-chiral fields. These are chiral or anti-chiral only on the left-moving or right-moving side and are denoted $(c,-)$ or $(a,-)$. These will be important in the discussion of the NSNS flux deformations in section \ref{sect:deformations}.

Elements of the chiral ring are classified by their $\U(1)_L\times\U(1)_R$
charges, and we have chosen the convention that chiral ring elements have
positive charge while anti-chiral ring elements carry negative charge. By
spectral flow, elements of the chiral ring can be mapped to Ramond ground
states. Furthermore, for a Calabi-Yau target space at large volume, chiral
ring elements map to harmonic forms on the CY. The ring will, in general,
be $\alpha'$ corrected athough it is expected to be robust under RG flow to
the IR fixed point. Elements of the $(c,a)$-ring with $\U(1)_L\times
\U(1)_R$ charges $(r,-s)$ can be mapped to elements of
$H^{r,s}_{\delbar}(M_6)$. Similarly, elements of $(c,c)$-ring with charges
$(r,s)$ can be mapped to elements of
$H^{(r,0)}_{\delbar}(M_6,\Lambda^{(0,s)}T)$ which are, in turn, associated
with harmonic $(r,3-s)$-forms via contraction with the anti-holomorphic
$(0,3)$-form $\bar\Omega$ of the Calabi-Yau. Chiral  and
twisted-chiral ring elements, as in RNS, correspond to complex structure and K\"ahler
deformations respectively.


At large volume, the operators in the chiral rings have a representation in terms of the Weyl fermions which makes manifest the relation between chiral ring elements and harmonic forms. We review this representation now as it will be used in section \ref{sect:deformations} to construct flux deformations.

All operators in the internal CFT are classified by their $\U(1)_L\times\U(1)_R$ charges. We denote an operator with charges $(p,q)$ by $\Omega_{p,q}$. Then a $(c,a)$-ring element $\Omega_{p,-q}$  with $\U(1)_L\times\U(1)_R$ charges $(p,-q)$ is (in the large radius limit):
\be
\Omega_{p,-q} = \omega_{i_1\ldots
i_p,\jbar_1\ldots\jbar_q}\psi_L^{i_1}\ldots\psi_L^{i_p}\bar \psi_R^{\jbar_1}\ldots\bar \psi_R^{\jbar_q},
\ee
where $\omega$ is a $(p,q)$-form. Physical state and chirality conditions imply $\omega \in H^{(p,q)}(\mathcal{M}_6)$. A $(c,c)$-ring element $\Omega_{p,q}$ corresponds to
\be
\Omega_{p,q} = \omega_{i_1\ldots i_p}^{\qquad \jbar_1\ldots \jbar_q}\psi_L^{i_1}\ldots\psi_L^{i_p}\psi_{R\jbar_1}\ldots\psi_{R\jbar_q},
\ee
which, after applying physical state conditions, gives $\omega \in H^{(p,0)}(M_6,T^{(0,q)}M_6)\cong H^{(p,3-q)}(M_6)$.


To construct a critical string worldsheet theory we take the direct sum of the space-time and $CY_3$ superconformal algebras. The left-moving generators become
\bea
G_L^+ &=& e^{\rho_L}(\bar d_L)^2 + \cG_L^+, \quad G_L^- = e^{-\rho_L}(d_L)^2 + \cG_L^-, \quad J_L = -\del_L \rho_L + \cJ_L, \label{eqn:algebra_6}
\eea
while the right movers are\footnote{This expression satisfies the type IIB GSO projection: The $4d$ space-time and internal CY chiralities are correlated. In type IIA they are anti-correlated.}
\bea
G_R^+ = e^{\rho_R}(\bar d_R)^2 + \cG_R^+, \quad G_R^- = e^{-\rho_R}(d_R)^2 + \cG_R^-, \quad J_R = -\del_L \rho_R + \cJ_R.\label{eqn:algebra_8}
\eea

\subsection{Physical States}
\label{sect:physical_states}
\subsubsection*{KK Reduction of Type IIB Supergravity on a $CY_3$}
The massless spectrum of type IIB string theory on a Calabi-Yau 3-fold consists of a gravity multiplet, a tensor multiplet, and massless moduli (coming from the K\"ahler and complex structure deformations). The supergravity multiplet has bosonic components
\be
G_{\mu\nu},B_{\mu\nu},\Phi,C_0,C_{\mu\nu},C_{\mu ijk}.
\ee
Additionally there are K\"ahler moduli  $s^a$ with $a=1,\ldots,h^{(1,1)}$ and complex structure moduli $\phi^p$ with $p=1,\ldots h^{(1,2)}$. The tensor fields are decomposed as
\bea
B_{MN} &=& \sum_a r^a \omega_a^{(1,1)},\cr
C_{MN} &=& \sum_a z^a \omega_a^{(1,1)},\cr
C_{MNPQ} &=& \sum_a v^a \omega^{(1,1)}_a\w \omega_a^{(1,1)} + \sum_p A^p_\mu \omega_p^{(1,2)}.
\label{eqn:KKcomponents}
\eea
This results in 4 real scalars $s^a,r^a,z^a,v^a$ for each $(1,1)$-form grouped into a $4d$, $N=2$ hypermultiplet. It is often convenient to write $\sigma^a = s^a + i r^a$ for a complexified K\"ahler modulus. There is also a $4d$, $N=2$ vector multiplet ($A_\mu^p$ and $\phi^p$) for each $(1,2)$-form. We will now construct the corresponding hybrid vertex operators.

\subsubsection*{Physical States in String Theory}
Physical states in RNS are Virasoro primaries satisfying
\be
L_n\cdot \cV = G_{r}\cdot \cV = 0\label{eqn:primary_1}
\ee
for $n,r>0$, modulo BRST exact states. The direct product structure \C{eqn:action_1} implies vertex operators are tensor products
\be
\cV = \Vext\otimes \Vint.
\ee
Elements of the gravity and tensor multiplets correspond to massless states with $\Vint=1$. The massless moduli correspond to the case when $\cV_{\rm int}$ is a chiral or twisted-chiral primary. The hybrid vertex operators are defined analogously except that $\Vext$ is constructed from 0-modes of $(x,\th,\thab)$. Consequently, physical states are automatically $4d$, $N=2$ superfields
\cite{Berkovits:1995cb}. The multiplets are:
\begin{enumerate}
\item {\it Gravity and tensor multiplets:}
$\cV = U \otimes 1$.

    These vertex operators are known as real primary fields. The physical state conditions require it to be an $N=2$ Virasoro primary and real $\cV^*=\cV$. It can be shown using the algebra \C{eqn:algebra_1}-\C{eqn:algebra_4} and \C{eqn:algebra_6}-\C{eqn:algebra_8} that this implies the polarization conditions and equation of motion
    \be
    \nabla_L^2 U = \bar\nabla_L^2 U = \nabla_R^2 U = \bar\nabla_R^2 U = \Box_4 U = 0,\label{eqn:u_phys_1}
    \ee
    where
    \be
    \nabla_{L\al}=\frac{\del}{\del\thal} + i(\sig^M\thab_{L})_{\al}\frac{\del}{\del X_M},\quad     \nabla_{R\al}=\frac{\del}{\del\thal} + i(\sig^M\thab_{R})_{\al}\frac{\del}{\del X_M}\label{eqn:CovDer}
    \ee
are the superspace covariant derivatives. The equations (\ref{eqn:u_phys_1})
are the linearized 1-loop $\beta$-function equations.\footnote{To be more
precise, we could interpret this as ``super-$\beta$-function equations''
\cite{skend,nedel} since they
come from the requirement that the superconformal generators still satisfy the
superconformal algebra after the action is deformed by this vertex
operator.} It can be shown that $U$ is a superfield containing the $N=2$ conformal supergravity muliplet and a tensor multiplet \cite{Berkovits:1995cb,Siegel:1995px}. The lowest component is given by
     \be
     U = (\thab_L \sigma^\mu\tha_L)(\thab_R\sigma^\nu\thab_R)(h_{\mu\nu} + b_{\mu\nu} + \eta_{\mu\nu}A) + \ldots\label{eqn:realu_1},
     \ee
     where $h_{\mu\nu}$ is the symmetric traceless metric, $b_{\mu\nu}$
the $B$-field and $A$ is related to the trace of the metric.\footnote{In
fact, the scalar field $A$ sits in the tensor multiplet. Only after fixing
space time conformal symmetry is this component related to the trace of
the metric.} The omitted fields are of higher dimension and will not be relevant for the discussions in this paper. The integrated vertex operator is
\bea
&&\int d^2z\, (G^+_L)_{-\half}(G^-_L)_{-\half} (G^+_R)_{-\half}(G^-_R)_{-\half} U.
 \eea
 Explicitly this becomes
 \bea
&& \int d^2 z \left\{   \cO_L \cdot \cO_R\cdot  U \right\}\non\label{eqn:real_integrated_1}
\eea
for
\bea
\cO_L &=&  d_{\alpha L}
\bar\nabla^2_L \nabla_L^\alpha -\bar d_{\dot\alpha L}
\nabla_{\dot\alpha L}\bar\nabla^2_L +\partial\theta^\alpha_L
\nabla_{\alpha L} - \partial \bar\theta^{\dot\alpha}_L
\bar\nabla_{\dot\alpha L}- i \Pi_{\alpha\dot\alpha}
[\nabla_L^\alpha,\bar\nabla_L^{\dot\alpha}] + \non\\ && {\cal G}_L^+\cdot {\cal
G}^-_L +
e^{\rho_L} \bar d_{L{\dot\alpha}}\bar\nabla^{\dot\alpha}_L {\cal G}^-_L +
e^{-\rho_L} d_{L\alpha}\nabla^\alpha_L {\cal G}^+_L  \non
\eea
(and the analogous expression for $\cO_R$), where all operators act on $U$.

Recall that the $\cG$s are the internal CY generators; in general we can allow $U$ to be a function of the coordinates of the internal manifold. However, this would imply that the last physical state condition in \C{eqn:u_phys_1} is not satisfied, since $U$ will have double poles with $\mathcal{T}$. We will return to this in section \ref{sect:applications} when we consider flux perturbation theory and the $\beta$-function. There we will see that this violation is of second order in the flux. This, in turn, implies that $U$ breaks conformal invariance at 1-loop in $\alpha'$.
We will ignore this issue for now and  return to it in section \ref{sect:applications}.
We conclude by noting that it is not possible for $U$ to depend on the fermions of the internal theory since this breaks conformal invariance already at classical level.

\item {\it Vector Multiplet}: $\cV_{CC} = \Vext \otimes \Vint$ with $\Vint \in (c,c)$-ring.

    In type IIB string theory, vector multiplets and chiral ring elements correspond to complex structure deformations. The physical state conditions impose that the vertex operator is a massless Virasoro primary \C{eqn:primary_1} and satisfies a chirality constraint:
    \bea
    (G_{L}^+)_{-\half} \cdot \cV_{CC} = 0,\quad (G_{R}^+)_{-\half} \cdot \cV_{CC} = 0.\label{eqn:chirality_1}
    \eea
    This implies that $\Vint \in (c,c)$-ring and $\Vext$ is a $4d$, $N=2$ chiral superfield which has the component expansion
    \bea
    \Vext &=& \phi + \theta_L^2 D_{--} + \theta_R^2 D_{++} + \theta_L\theta_R D_{+-} + \theta_L \sigma^{\mu\nu}\theta_R F_{\mu\nu} + \theta_L^2 \theta_R^2 \Box \bar \phi\non\\&&+{~\rm fermions}\label{eqn:M_CC_1},
    \eea
    where $\phi$ is the complex structure modulus. The $D_{ij}$ are auxiliary fields with $i,j$ $\SU(2)$ $R$-symmetry indices, while $F_{\mu\nu}$ is the field strength for the vector field. This is the field content of the $N=2$ vector multiplet. The physical state conditions \C{eqn:primary_1} put \C{eqn:M_CC_1} on-shell
    \be
     \nabla_L^2 \Vext(x,\tha_L,\tha_R) =  \nabla_R^2 {\Vext}(x,\bar \tha_L,\bar \tha_R) = 0\label{eqn:eom_1}
    \ee
    which removes the auxiliary fields and imposes the equations of motion on the physical field content. 
    The integrated vertex operator is of the form
    \be
    \int d^2 z G_{L,-\half}^- G_{R,-\half}^- \cV_{CC} + \int d^2 z G_{L,-\half}^+ G_{R,-\half}^+ \ol{\cV}_{CC},
    \label{eqn:CC_integrated_1}
    \ee
   where the complex conjugate is included for unitarity.

\item {\it Hypermultiplets}: $\cV = \Vext \otimes \Vint$ with $\Vint \in (c,a)$-ring.

    The hypermultiplets in type IIB correspond to K\"ahler deformations and are twisted-chiral ring elements. The vertex operator is a massless Virasoro primary satisfying \C{eqn:primary_1} and a twisted chirality constraint:
    \bea
    (G_{L}^+)_{-\half} \cdot \cV_{CA} &=& 0,\quad (G_{R}^-)_{-\half} \cdot
\cV_{CA} = 0.\label{eqn:twisted-chiral_1}
    \eea
    These conditions imply that $\Vint \in (c,a)$-ring. The corresponding $4d$ superfield has a theta expansion
    \bea
     \Vext &=& l_{++} + \theta_L^2 y + \bar\theta_R^2 \bar y + \theta_L \sigma^{\mu}\bar \theta_R (\del_{\mu} l_{+-} + (*H)_\mu) + \theta_L^2 \bar \theta_R^2 \Box l_{--}\non\\&& +{\rm ~fermions},\label{eqn:M_AC_1}
    \eea
    where
    $l_{ij}$ are scalars in the tensor representation of $\SU(2)_R$,  $y,\bar y$ are auxiliary fields and $H_{\mu\nu\lambda}$ is a tensor field strength. This is the field content of the $N=2$ hypermultiplet. The primary conditions
    \be
     \nabla_L^2 \Vext(x,\tha_L,\tha_R) =  \ol{\nabla}_R^2 {\Vext}(x,\bar \tha_L,\bar \tha_R) = 0.\label{eqn:eom_2}
     \ee
     put the field content on-shell.
   In components these equations fix the auxiliary fields and imply the equations of motion and polarization conditions for the remaining component fields. The integrated vertex operator is of the form
    \be
    \int d^2 z (G^-_L)_{-\half} (G^+_R)_{-\half} \cV_{CA} + \int d^2 z
(G^+_L)_{-\half} (G^-_R)_{-\half} \ol{\cV}_{CA}.
    \label{eqn:CA_integrated_1}
    \ee
\end{enumerate}

With the (space-time)$\otimes$(internal) structure of the vertex operators understood, we will drop the cumbersome tensor product notation henceforth.

\section{Flux Deformations}
\label{sect:deformations}
In the previous section, we listed the physical states of string theory as hybrid vertex operators. As in RNS, these vertex operators can also be used to construct deformations of theory. For example, one can construct integrated vertex operators corresponding to complex structure and K\"ahler deformations in the obvious way as
\be
\int d^2 z (G^-_L)_{-\half}(G^-_R)_{-\half} \cdot \left( \phi_p \Omega^p_{(1,1)} \right)
\quad \mathrm{and}
 \quad \int d^2 z (G^-_L)_{-\half}(G^+_R)_{-\half} \cdot \left( (l_{++})_{k} \Omega^k_{1,-1} \right),
\ee
where $p=1,\ldots, h^{(1,2)}$ and $k=1,\ldots , h^{(1,1)}$ label a basis of $(c,c)$ and $(c,a)$ operators of the internal CFT respectively. 

In addition to the massless moduli, there are also RR and NSNS fluxes. The fluxes of most interest to us are the 3-form fluxes on the internal space. In supergravity these may be decomposed as
\bea
F_{MNP} &=& \sum_p F_{(1,2)}^p \omega_p^{(1,2)} + \cc\cr
H_{MNP} &=& \sum_p H_{(1,2)}^p \omega_p^{(1,2)} + \cc\label{eqn:fluxes_1}
\eea
The vertex operators for these fluxes are well-known in RNS, and in particular, the RR vertex operators have branch cuts in their operator products. The remarkable  property of the hybrid is that even though the internal variables are represented in the RNS formalism, the manifest space-time supersymmetry removes the branch cuts complicating the space-time part of that formalism. The resulting GSO projected RR flux vertex operators may consistently be integrated into the action. 
In this section we will construct the flux vertex operators.

Our starting point will be the RNS formalism. The space-time supercharges are given in the $-\half$-picture by
\bea
Q_{L\al} &=& \oint dz e^{-\half \phi_L}\Sig_{L\al} e^{-i \sqrt{3} H_{L}/2}, \quad
\bar Q_{L\da} = \oint dz e^{-\half \phi_L}\bar \Sig_{L\da} e^{i\sqrt{3}H_L/2},\cr
Q_{R\al} &=& \oint d\bz e^{-\half \phi_R} \Sig_{R\al} e^{-i \sqrt{3} H_{R}/2},\quad
{\bar Q}_{R\da} = \oint d\bz e^{-\half \phi_R}{\bar \Sig}_{R\da}  e^{i\sqrt{3} H_{R}/2},\non\\\label{eqn:susy_1}
\eea
where $\Sig_L,\Sig_R$ are the left- and right-moving spin fields and $e^{\pm i\sqrt{3}H_{L,R}/2}$ is the operator implementing spectral flow by half a unit. The GSO projection has been applied, that is, all four $Q$'s above have eigenvalues $-1$ with respect to $e^{i\pi F}$.

Now consider the RNS vertex operator for the complex structure modulus $\phi^p$, which is an element of the $(c,c)$-ring. That is,
\be
k^2 \phi^p = e^{-\phi_L-\phi_R}  e^{ikX} \Omega^p_{(1,1)}, \label{eqn:cs_1}
\ee
where $\Omega_{(1,1)}^p = \omega_{i}^{~\bar j}(X) \psi_L^i \psi_{R,\jbar}$ and $\omega \in H^{1}(M,TM)$. We will now use equation \C{eqn:cs_1} and the $N=2$ space-time supersymmetry algebra of the superchrages  \C{eqn:susy_1} to identify the vertex operators with the auxiliary fields $D_{ij}$ and hence fluxes in the hybrid. We follow the techniques developed in \cite{Atick:1987gy,Dine:1987gj}. Some recent analysis along these lines can be found in \cite{Lawrence:2004zk,Lawrence:2007jb}.

\subsection{$F_3$ Flux}
\label{sect:f3}
The RR vertex operator for $F_3$ flux on the internal space is constructed as a spinor bilinear in the RNS spin-fields. One can check that the correct vertex operator is
\be
 e^{-\half \phi_L - \half \phi_R} \eps^{\al\bet}\Sig_{L\al} \Sig_{R\bet} \Psi_{RR}^{p},\label{eqn:rr_1}
\ee
where $\al,\beta$ are $4d$ spinor indices and $\Psi_{RR}^{p}$ is the RR ground state in the internal CFT corresponding to the $(1,2)$-cohomology representative.
From the space-time supersymmetry algebra we have $k^2 \phi^p = \eps^{\dal\db} \{ {\bar Q}_{L,\dal}, [{\bar Q}_{R,\db},D^p_{+-}]\}$. Following \cite{Atick:1987gy,Dine:1987gj,Lawrence:2004zk,Lawrence:2007jb} and using \C{eqn:susy_1} and \C{eqn:cs_1} we can identify
\be
D^p_{+-} \leftrightarrow i g_s F_{(1,2)}^p \leftrightarrow g_s e^{-\half \phi_L - \half \phi_R} \eps^{\al\bet}\Sig_{L,\al} \Sig_{R,\bet} \Psi_{RR}^{p}.\label{eqn:d_rr_1}
\ee
This identification agrees with the analysis in \cite{Lawrence:2004zk}. Using the component expansion \C{eqn:M_CC_1} and the correspondence \C{eqn:d_rr_1} we can identify the hybrid vertex operator corresponding to $F_{(1,2)}$ flux as
\be
\cV_{F_{(1,2)}} = ig_s \tha_L\tha_R F_{(1,2)}^p\Omega^p_{(1,1)}.
\ee
This satisfies the physical state conditions \C{eqn:primary_1} and is a chiral ring element in the full $d=10$ hybrid theory. It is now possible to see, schematically at least, how the RNS to hybrid map is working:
 \begin{itemize}
 \item The spin fields and ghosts map as $e^{-\phi_L -{\phi_R}} \e^{\al\beta} \Sigma_{L\al}\Sigma_{R\beta} \leftrightarrow \tha_L \tha_R$.
     \item The RR ground state is transformed by spectral flow to the NSNS sector: $\Psi^p_{RR} \leftrightarrow \Omega_{(1,1)}^{p}$.
 \end{itemize}
 As the $\tha_{L,R}$ variables have worldsheet spin $0$, they have no branch cuts in their operator products. Using the form of the integrated vertex operator \C{eqn:CC_integrated_1} we can write down the explicit deformation of the action:
\bea
\delta S_{F_3} &=& i g_sF^{(1,2)}_p \int d^2z \left\{ e^{-\rho_L-\rho_R} d_L d_R \Omega_{(1,1)}^p + (\theta_L\theta_R) (\mathcal{G}_L^-  \mathcal{G}_R^- \cdot \Omega_{(1,1)}^p) +\right.\cr
 && + \left.e^{-\rho_L}(d_L\tha_R) (\cG_{R,-\half}^-\Omega_{(1,1)}) + e^{-\rho_R}(d_R\tha_L) (\cG_{L,-\half}^-\Omega_{(1,1)})\right\} + \cc\non\\
\eea
The complex conjugate corresponds to $F^p_{(2,1)}$ fluxes in the obvious way and is required for unitarity.

\subsection{$H_3$ Flux}
\label{sect:h3}
In the RNS $\sigma$-model describing a Calabi-Yau background, a deformation of the worldsheet action by a certain $H_3$ flux corresponds to
\bea
\delta S &=& \frac{1}{4\pi} \int d^2 z \left\{\,(\delta G_{i \bar j}(X) + \delta B_{i \bar j}(X))\del_L X^i\del_R X^{\bar j} + \left(\delta \Gamma_{\ibar\jbar k}(X) +  \half \delta H_{\ibar \jbar k}(X)\right)\bar\psi_L^\ibar  \bar\psi_L^\jbar \del_R X^{ k}\right.\cr
&& \left.  +\left(\delta \Gamma_{\ibar\jbar k}(X) - \half \delta H_{\ibar\jbar k}(X)\right)\bar\psi_R^\ibar \bar\psi_R^\jbar \del_L X^{k} + {\rm c.c.} + \ldots\right\}\label{eqn:b_def_1},
\eea
where all quantities are small deformations of the $N=(2,2)$ Calabi-Yau background: $\delta B$ is the $B$-field deformation and $\delta G$ is the associated metric deformation required to preserve supersymmetry. $\delta \Gamma$ is the deformation of the Levi-Civita connection due to $\delta G$ while $\delta H$ is the field strength corresponding to $\delta B$. The omitted terms are the 4-fermion terms which will not be relevant for the discussion in this section.  All of these terms are required to preserve worldsheet supersymmetry.  For notational convenience we drop the $\delta$ henceforth, keeping in mind that these operators are to be regarded as small deformations.

We now turn to the construction of the analogue of the RNS deformation \C{eqn:b_def_1} in the hybrid formalism.  First, we construct the vertex operators for the torsional connection $\Gamma \pm \half H$. Then, we identify the vertex operator for the complexified metric and point out that only the total vertex operator is required to be physical. This resolves the issue discussed in \cite{Berkovits:2003pq,Lawrence:2004zk,Linch:2006ig} of how to consistently turn on fluxes on the internal space.

The $N=2$ supersymmetry algebra implies that the $D_{++}$ auxiliary component field satisfies
\be
\eps^{\dal\db} \{ {\bar Q}_{R\dal}, [ {\bar Q}_{R\db},D^p_{++}]\} = k^2 \phi^p.
\ee
Using the same procedure as that above, we find the corresponding RNS vertex operator
\be
D^p_{++} \leftrightarrow \omega^p_{i \jbar \kbar}  \del_L X^{i}\bar \psi_R^{\jbar} \bar \psi_R^{\ibar}.\label{eqn:d++_1}
\ee
Here, $\omega^p \in H^{1,2}(M)$ and we have applied picture changing so
that the vertex operator is in the $(0,0)$-picture. From the deformation
\C{eqn:b_def_1}, we read off that the term $\bar \psi_R \bar \psi_R \del_L X$ couples
to a combination of the Christoffel connection $\Gamma^M_{\,NP}$ and the
$B$-field. Thus, we identify\footnote{We deduce the presence of the $i$
either through  the corresponding vertex operator for the metric or the $B$-field.}
\be
D^p_{++}  \leftrightarrow i \Gamma^p_{(1,2)} + \frac{i}{2} H_{(1,2)}^p.\label{eqn:d++_2}
\ee
Similarly, by interchanging left-movers and right-movers, one identifies\footnote{This identification differs from the results of the analysis in \cite{Lawrence:2004zk} in which it was claimed that $D^p_{\pm\pm}\leftrightarrow T^p_{(1,2)} \pm H^p_{(1,2)}$ with $T$ is the (metric) torsion tensor.
}
\be
D^p_{--}  \leftrightarrow i \Gamma^p_{(1,2)} - \frac{i}{2} H_{(1,2)}^p.\label{eqn:d--_1}
\ee
With this, the hybrid vertex operator corresponding to the torsional connection in \C{eqn:b_def_1} becomes
\bea
\cV_{H_3} \sim i\left(\Gamma^p_{(1,2)} - \frac{1}{2} H_{(1,2)}^p\right)\th_L^2 \Omega^p_{(1,1)} + i\left(\Gamma^p_{(1,2)} + \frac{1}{2} H_{(1,2)}^p\right)\th_R^2 \Omega^p_{(1,1)} .\label{eqn:hybrid_h_3}
\eea

As it stands, this vertex operator is not a Virasoro primary since it has a non-zero double-pole with $G_{L,R}^-$. Equivalently, this vertex operator does not satisfy the physical state conditions. This implies that we cannot turn on this vertex operator since it corresponds to auxiliary fields which are forced to vanish by the equations of motion. This is the puzzle raised in \cite{Berkovits:2003pq,Lawrence:2004zk}, a solution to which was proposed in \cite{Linch:2006ig}: One should add other terms violating the physical state conditions in such a way that the sum does not, and consequently, is physical. From the explicit formula for the deformation (\ref{eqn:b_def_1}) above, we can now give a physical interpretation to these pieces: The vertex operator (\ref{eqn:hybrid_h_3}) is missing the metric and $B$-field.

Motivated by \cite{Linch:2006ig} we are led to propose the following vertex operator for the NSNS flux:
\be
\cV_{H_3} = i\left(\Gamma^p_{(1,2)} - \frac{1}{2} H_{(1,2)}^p\right)\th_L^2 \Omega^p_{(1,1)} + i\left(\Gamma^p_{(1,2)} + \frac{1}{2} H_{(1,2)}^p\right)\th_R^2 \Omega^p_{(1,1)} + e^{-\rho_L}E_{(2,1)} + e^{-\rho_R}E_{(1,2)}, \label{eqn:def_2_c}
\ee
where the operators $E_{(i,j)}$ have $\U(1)$ charges $(i,j)$. Let us now interpret the last two terms. We require this operator to be physical in order to preserve $N=(2,2)$ worldsheet supersymmetry. The physical state conditions imply for the left-movers that
\be
\cG_{L,\half}^- \cdot E_{(2,1)} = i\left(\Gamma^p_{(1,2)} - \frac{1}{2} H_{(1,2)}^p\right)\Omega^p_{(1,1)}.\label{eqn:e_1}
\ee
The operator $E_{(2,1)}$ is, therefore, not in the chiral ring. To interpret it we can go to the large radius limit, where
\be
E_{(2,1)} = E_{ij}^{~~\bar k} \psi_L^i \psi_L^j \psi_{R,\bar k} = \Omega^+_{L} (E_{\bar i j} \bar \psi_L^{\bar i} \psi_R^j),
\ee
and $\Omega^+_L = \Omega_{ijk}\psi_L^i\psi_L^j\psi_L^k$ is the operator corresponding to the holomorphic 3-form. Then,
\be
\cG_{L,\half}^- \cdot E_{(2,1)} = \Omega_{L}^+ (-i\delbar_{\ibar } E_{\jbar k} \bar \psi_L^\ibar \bar\psi_L^\jbar \psi_R^k).\label{eqn:e_3}
\ee
Noting that
\bea
\left(\Gamma^p_{\ibar\jbar k} - \frac{1}{2} H_{\ibar\jbar k}^p\right) &=& -i(\delbar_{[\ibar}G_{\jbar]k} + \delbar_{[\ibar} B_{\jbar] k})\label{eqn:e_4},
\eea
and plugging this and \C{eqn:e_3} into \C{eqn:e_1}, we find that\footnote{The operator $e^{-\r_L}\Omega_L^+$ is an $\SU(2)_R$ rotation  which arises when one embeds the critical $N=2$ string as a topological $N=4$ theory. See for example \cite{Berkovits:1994vy}.}
\be
 E_{2,1} = -i\Omega_L^+ (G_{\ibar j} + B_{\ibar j}) \bar\psi_L^{\ibar} \psi_R^j.
\ee
Thus, this new term corresponds to the complexified metric. The right-movers follow analogously. By construction, the full vertex operator \C{eqn:def_2_c} satisfies the physical state conditions. It may be used to deform the action by the formula for the integrated operator \C{eqn:CC_integrated_1}.

We close this section with the observation that the operators $B_{(2,1)}$ and $B_{(1,2)}$ are semi-chiral fields. This implies that the hybrid description of NSNS flux backgrounds is intimately related to the torsional $\sigma$-models discussed in \cite{Lindstrom:2005zr,Lindstrom:2007vc} and references therein. 
Direct investigation of the interplay between generalized geometry and target space supersymmetry should be possible using this observation.

\subsection{$G_3$ Flux}
Using the results of sections \ref{sect:f3} and \ref{sect:h3} it is now
straightforward to write down the vertex operator corresponding to $G_3 =
F_3 - \tau H_3$, which is the relevant flux for conformally Calabi-Yau
compactifications. We will set the Ramond-Ramond scalar $C_0=0$ here for simplicity,

It will be convenient to perform a field redefinition on the vector multiplet components \C{eqn:M_CC_1} so that
\begin{eqnarray}
\Vext &=& \phi^p
+(\theta^-)^2X^p_{--}+(\theta^+)^2X^p_{++}+ \theta^-\theta^+ X^p_{+-}+\dots,
\end{eqnarray}
where $ \theta^\pm = \theta_L \pm i\theta_R$, and the new fields are
\bea
4 X^p_{++}&=&D^p_{++}-D^p_{--}-iD^p_{+-},\cr
4 X^p_{--}&=&D^p_{++}-D^p_{--}+iD^p_{+-},\cr
2 X^p_{+-}&=&D^p_{++}+D^p_{--}.\label{eqn:relations_1}
\eea
The identifications of the vector multiplet auxiliary fields \C{eqn:d_rr_1}, \C{eqn:d++_2} and \C{eqn:d--_1} give
\be
4X_{++} = g_s (F_{(1,2)}^p + ig_s^{-1}H_{(1,2)}^p) = g_s { \ol{G}_{(2,1)}}, \quad 4X_{--} = g_s (F_{(1,2)}^p - ig_s^{-1}H_{(1,2)}^p)= g_s G_{(1,2)}\label{eqn:x_1}.
\ee
Suppose we want to turn on $G_{(2,1)}$ flux and keep $G_{(1,2)} = 0$. This corresponds to the vertex operator
\be
\cV_{G_{(2,1)}} = (\bar \th_L - i\bar \th_R)^2 g_s G^p_{(2,1)}\Omega^p_{(-1,-1)}, \quad \cV_{\ol{G}_{(2,1)}} = (\th_L + i\th_R)^2 g_s \ol{G^p}_{(2,1)}\Omega^p_{(1,1)},\label{eqn:g_21}
\ee
where $G_{(2,1)}^p$ is the amount of $G_3$ flux through the homology cycle labelled by $p=1,\ldots, h^{(2,1)}$. The integrated vertex operator is easily constructed using the rule \C{eqn:CC_integrated_1}.

Let us pause for a moment to make some comments. Firstly, giving an
expectation value to an auxiliary field multiplying $\theta^2$ breaks the corresponding supersymmetry. For example the vertex operator $\cV \propto (\th_L+i\th_R)^2$ breaks the corresponding space-time supersymmetry generated by $(Q_L + i Q_R)$. Secondly, we see from the explicit form \C{eqn:g_21} of the $G_{(2,1)}$ vertex operator that turning on $(2,1)$-flux is compatible with the $N=1$ supersymmetry preserved by a D3-brane, as expected. Conversely, $G_{(1,2)}$ corresponds to anti-D3-branes, which preserve the opposite supersymmetry. It is also obvious here that turning on both $G_{(2,1)}$ and $G_{(1,2)}$ breaks all space-time supersymmetry as would be expected by adding D3-branes and anti-D3-branes. This is entirely in agreement with the supergravity literature \cite{Grana:2001xn,Grana:2005jc}, but such features would be difficult to see from the worldsheet in the RNS description since the space-time supersymmetry is not manifest.

\subsection{$F_1$ and $F_5$ Flux}
\label{sect:f1_f5}
It is straightforward to follow the above reasoning to deduce the form for the vertex operators corresponding to $F_1$ and $F_5$ fluxes where the $F_5$ is space-time filling and $F_1$ lies in the internal space. The result is that the vertex operators are given by
\bea
F_1 &\rightarrow& i F_1 \th_L \th_R \Omega_{(1,0)}, \\
F_5 &\rightarrow& F_5 \th_L \th_R \Omega_{(1,0)}.
\eea
For compact Calabi-Yau 3-folds with exactly $\SU(3)$ holonomy and not a proper subgroup, the cohomology in dimension 1 vanishes: $h^{1,0}=h^{0,1}=0$. Nevertheless, it is interesting to study these vertex operators for two reasons. Firstly, $F_5$ appears in conformally Calabi-Yau compactifications through a
Bianchi identity, so it cannot be ignored in such backgrounds. Secondly, these flux operators
appear in compactifications preserving $16$ and $32$ supersymmetries ({e.g.}~compactifications on $M_6=T^6$ preserve $32$ supersymmetries). These extra supersymmetries are non-linearly realized in the hybrid formalism. The $F_5$ and $F_1$ strengths appear in multiplets which contain fields that are related to these non-linearly realized supersymmetries. Here we
will describe them in the simplest possible case of compactification
on $T^6$.

The multiplets in question are semi-chiral on the worldsheet. They have conformal weights $(\half,0)$
and $(0,\half)$ and take the form
\begin{equation}
S_L = s_{i\, L} \psi_L^i,\quad \bar S_L =\bar s_{\ibar\, L}
\bar\psi_L^{\bar\imath},\quad S_R = s_{i\, R}\psi_R^i,\quad \bar S_R=\bar
s_{\ibar\, R} \bar\psi_R^{\bar\imath}.
\end{equation}
The physical state conditions for $S_L$ are
\begin{equation}
G^+_{L,-\half} S_L = G^-_{L,\half} S_L =
G^+_{R,\half}S_L=G^-_{R,\half}S_L=0,
\end{equation}
with analogous conditions for $\bar S_L$, $S_R$, and $\bar S_R$. In terms
of the space-time fields $s_{i\, L}$ these conditions are
\begin{equation}
\bar\nabla_{\dot\alpha\, L} s_{i\, L} = \nabla^2_L s_{i\,
L}=\bar\nabla^2_R s_{i\, L}= \nabla^2_R s_{i\, L}=0.
\end{equation}
In the integrated vertex operator the semi-chiral space-time superfield admits a gauge transformation
\begin{eqnarray}
\delta s_{iL}=\nabla_R^2 \Xi_{iL} + \bar \nabla_R^2 \Upsilon_{iL}
\end{eqnarray}
with semi-chiral parameter fields $\Xi$ and $\Upsilon$. These can be used to put $s_{iL}$ into the Wess-Zumino gauge
\begin{equation}\label{spin32}
s_{i\, L}= \theta_R \sigma^\mu \bar\theta_R B_{\mu\, i} +
\theta_R\sigma_\mu\bar\theta_R \theta_L^\alpha( \psi^\mu_{\alpha\, i} +
\sigma^\mu_{\alpha\dot\alpha} \chi^{\dot\alpha}_i ) +
\theta_L\theta_R\bar\theta^2_R (iF_{5\, i}+F_{1\, i})+\dots,
\end{equation}
where we see that the $F_1$ and $F_5$ fluxes form a complex auxiliary scalar
inside $s_{i\, L}$. A similar analysis holds for $S_R$, $\bar S_L$ and
$\bar S_R$. We will elucidate the role played by these auxiliary fields in section \ref{CCYBG}.

\section{Physical Effects and Applications}
\label{sect:applications}
In this section we study the consequences of turning on $G_3$ flux using the identifications made above. First, the supergravity description of the class of flux backgrounds we wish to study is reviewed. The same backgrounds are then studied from the string worldsheet and we demonstrate precise agreement with supergravity.

\subsection{Conformally $CY_3$ Backgrounds in Supergravity}
\label{sect:sugra}
The class of flux solutions that we consider in this paper are well-studied in the supergravity literature. They are known as ``conformally Calabi-Yau'' solutions and are one of the simplest classes of flux solutions. Examples of this type were first constructed in \cite{Dasgupta:1999ss}, and a review with references is in \cite{Grana:2005jc}. We consider such backgrounds in order to compare our string calculations with known results in a regime where they should agree, that is, at large volume. Let us now review some of the relevant aspects of such compactifications and give a simple example we can study.

The geometry is warped, with metric given by
\be
ds^2 = e^{2A(y)} \eta_{\mu\nu} dx^\mu dx^\nu + e^{-2A(y)} g_{ij} dy^i dy^j,
\ee
where $e^{2A(y)}$ is the warp factor. The internal space is related to a Calabi-Yau 3-fold $M_6$ with metric $g_{ij}$ by a conformal factor $e^{-2A(y)}$. There is three-form flux $G_3 = F_3 - \tau H_3$ which, in order to be compatible with $N=1$ space-time supersymmetry, is taken to be a $(2,1)$-form:\footnote{Typically in the literature, the $G_3$ flux is taken to be of type $(2,1)$ and imaginary self-dual.
Space-time supersymmetry can also be achieved by choosing the
flux to be $(1,2)$ and imaginary anti-self-dual.} $G_3 \in
H^{(2,1)}(M_6)$. The equations of motion are satisfied if the flux is
primitive and imaginary self-dual. Additionally, there is a space-time filling 5-form flux which is proportional to derivatives of the
warp-factor. The fluxes also need to satisfy the appropriate
Bianchi-identities. The warp factor satisfies a Poisson equation sourced by
the 3-form fluxes
\be
\Box_{CY} e^{4A} =  \qrt g_s^2 |G_3|^2 + \ldots,\label{eqn:warp_1}
\ee
where the omitted terms are higher order in $\alpha'$. These may correspond to D-branes or  orientifold planes. In compact models, the presence of the orientifold planes is required in order for there to be non-trivial solutions to the Poisson equation \C{eqn:warp_1}. Further, one can generalize these solutions to include a holomorphically varying axio-dilaton $\tau$ which is sourced by the D7/O7-branes. For simplicity we will not consider any localized sources (either D3-branes, or $7$-branes) and assume that $C_0=0$ and that the dilaton is constant and tunable.

In this paper, we restrict ourselves to studying flux perturbation theory. The string theory analysis is then simplified and we find that we can reproduce known results from supergravity. To this end, consider a Calabi-Yau 3-fold with non-compact $(2,1)$-cycles. Fluxes supported by these cycles are not quantized. In this situation one can turn on small amounts of $G_3$ flux and study the physical consequences of such a deformation order by order in the flux.

To simplify the comparison with most of the literature, we require our flux to be compatible with D3-branes (as opposed to anti-D3-branes). The corresponding supersymmetry conditions imply that the $G_3$ flux is of type $(2,1)$.
This condition is known to lift complex structure moduli and can be represented by a superpotential
\be
W = \int G_3 \w \Omega.
\ee
in the low energy effective action \cite{Gukov:1999ya,Taylor:1999ii}.

At second order in the flux, the space is deformed so as to satisfy the equations of motion: A non-trivial warp-factor satisfying equation \C{eqn:warp_1} is generated. As we are doing perturbation theory in the flux, we can expand $e^{4A} = 1 + 4A + \ldots$, with the omitted terms being of higher order. There is also the 5-form flux, which is required for the supergravity equations of motion to be satisfied. It takes the form
$F_{0123i} = -4 \del_i A$. The Bianchi identities need to be satisfied, which for $G_3$ is trivial and for $F_5$ is equivalent to equation \C{eqn:warp_1}. We will now study such backgrounds from the string worldsheet and exactly re-derive these effects as a result of the deformations by the flux vertex operators.

\subsection{Conformally $CY_3$ Backgrounds in String Theory}
\label{CCYBG}

We now consider the class of backgrounds discussed in the previous section, that is, we turn on a small amount of $G_3$ flux along non-compactly supported cycles of a Calabi-Yau 3-fold. This corresponds to a deformation given by the integrated version
\bea
\cO_{G_3} &=& (G^-_L)_{-\half} (G^-_R)_{-\half}  \left[(\th_L + i\th_R)^2 g_s \ol{G^p}_{(2,1)}\Omega^p_{(1,1)}\right]  \cr
&&\qquad + (G^+_L)_{-\half} (G^+_R)_{,-\half}\left[(\bar \th_L - i\bar \th_R)^2 g_s G^p_{(2,1)}\Omega^p_{(-1,-1)}\right].
\eea
 of the vertex operator \C{eqn:g_21}. We now check conformal invariance order by order in $\alpha'$
and find that there are corrections to the background at string tree level that are required to maintain the conformal invariance (see, for example, \cite{Berenstein:1999ip} for a similar discussion in context of RNS). These corrections will correspond to space-time warping. 

\subsubsection*{Classical Superconformal Invariance}
The classical conditions for superconformal invariance were discussed in section \ref{sect:deformations}. There we found the following relevant features:
\begin{itemize}
\item The flux is required to be of type either $(2,1)$ or $(1,2)$. This is consistent with backgrounds compatible with either D3-branes or anti-D3-branes.
\item One cannot add $H_3$ independently of the explicit $B$-field terms, which come from semi-chiral fields.
\end{itemize}
\subsubsection*{1-loop Superconformal Invariance and Warping}
\label{sect:1-loop}
The 1-loop $\beta$-function is well-known to be equivalent to the space-time fields obeying their equations of motion to lowest order in $\alpha'$. The 1-loop $\beta$-function can be calculated by looking at the UV structure of the two-point function of two integrated vertex operators:
\bea
&&\Bigl< ... \int d^2z d^2w \cO_{G_3}(z) \cO_{G_3}(w)\Bigr> \sim \int
d^2zd^2w \frac{ \Pi^m_L \Pi_{m R}\,g_s^2 G_{(2,1)}^p
\ol{G}_{(2,1)}^{\bbar} g_{a\bbar},}{|z-w|^4} +
\ldots,\non\\\label{eqn:twopt_1}
\eea
where the ellipsis denotes arbitrary operator insertations and $g_{a\bbar}$ is the Zamolodchikov metric on the complex structure moduli space. The latter is defined by the 2-point function of Calabi-Yau chiral ring operators:
\be
\langle\Omega_{(1,1)}^p(z,\bz) \Ombar_{(-1,-1)}^\qbar(0,0)\rangle =\frac{g^{p\qbar}}{|z|^4}.
\ee
The contribution \C{eqn:twopt_1} induces a divergence
\be-\log(\Lambda)\int d^2z \Pi^m_L\Pi_{m R} g_s^2|G_{(2,1)}|^2\ee
which breaks conformal invariance. Conformal invariance can be preserved by introducing a correction to the background. Inspection of the component form of the physical states in section \ref{sect:physical_states} reveals that the only available candidate that preserves $4d$ Poincare invariance is the real primary vertex operator in \C{eqn:realu_1} with
\be\label{deltau}
U = (\th_L \sig^\mu \thab_L) (\tha_R \sigma^\nu \thab_R)\, \eta_{\mu\nu} A(y)+ \ldots,
\ee
where the scalar $A(y)$ is a function of the internal space. In the string gauge
this component field becomes the trace of the metric.

One of the contributions of the operator (\ref{deltau}) to the action is
\be\delta S_2=\int d^2z \Pi^m_L\Pi_{m R} A(y) +\dots\ee
When computing the contribution of the above term to the $\beta$-function
one can use the covariant background field expansion, which gives
a term
\be \int d^2z \Pi^m_L\Pi_{mR}  y^i  \bar
y^{\ibar} \nabla_i \bar\nabla_{\ibar} A(y),\ee
which, at 1-loop, gives a divergent term
\be\log(\Lambda) \int d^2z \Pi^m_L\Pi_{m R} \Box_{CY} A.\ee
The divergence is therefore canceled if the metric obeys the equation of motion
\be
\Box_{CY} A = g_s^2 |G_{(2,1)}|^2 + \ldots,
\ee
where $\Box_{CY}$ is the Laplacian on the Calabi-Yau. We have thus succeeded in
deriving the warping of space-time
from the worldsheet and it agrees precisely with the equation \C{eqn:warp_1} derived in
supergravity to this order in the flux.\footnote{There are additional
terms contained in the ellipsis from higher order calculations which give
higher order corrections to the background. }

We have not yet needed
to use the space-time-filling 5-form. Based on the supergravity analysis
in section \ref{sect:sugra}, this should appear quadratically in the flux.
It turns out that this space-filling 5-form flux appears in exactly the
same way as the warping, only in a different superfield. The easiest
way to see this is from the unintegrated $G_{(2,1)}$ flux vertex operator
\C{eqn:g_21}. Besides the divergence generating the warp factor, at
1-loop we get a divergent contribution from
\begin{equation}\label{div}
{\cal V}_{G_{(2,1)}}(z){\cal V}_{\bar G_{(2,1)}}(z)\sim
\frac{g_s^2}{\epsilon^2}(\bar\theta_L-i\bar\theta_R)^2(\theta_L+i\theta_R)^2
|G_{(2,1)}|^2
\end{equation}
giving rise to contributions with various $\theta$ dependence.
One of them is just the one which gives the warp factor but there are
others, for example, the $ -2i\theta_L\theta_R \bar\theta^2_R$ term. This
is going to be canceled by the last term in the component expansion \C{spin32} for the semi-chiral field. To be
more precise, if $S_L = i\theta_L\theta_R\bar\theta^2_R F_{5\, i}
\psi^i_L$, the contribution to the action has a term
\begin{equation}
({G}^-_L)_{-\half}({G}^+_R)_{-\half}({G}^-_R)_{-\half}S_L=\cdots + i\theta_L\theta_R\bar\theta^2_R
\bar\nabla_{\jbar}F_{5\, i}\partial_L y^i \partial_R \bar y^{\jbar}
+\cdots,
\end{equation}
and, if we regularize the product $\partial_L y^i \partial_R \bar y^{\jbar}$
using $ y^i(z)\bar y^{\jbar}(w)\to g^{i\jbar}\log( |z-w|^2+\epsilon^2)$, we get
\begin{equation}
\langle\partial_L y^i \partial_R \bar y^{\jbar}\rangle= -\frac{g^{i\jbar}}{\epsilon^2}.
\end{equation}
The divergence coming from the $G_{(2,1)}$ flux is canceled provided
\begin{equation}
\bar\nabla^i F_{5\, i} = 2g_s^2 | G_{(2,1)}|^2,
\end{equation}
and this gives the desired answer. All other terms coming from the operator product \C{div}
cancel in a similar way.

We emphasize that what were separate conformal field theories for the $4d$ space-time and internal Calabi-Yau have become mixed into a single conformal field theory for the 10-dimensional target space as a whole.
This is reflected in the warping of the $4d$ space-time and is expected to be a generic feature of flux backgrounds.

\subsection{Superpotential For Moduli}
\label{sect:superpontetial}
The presence of fluxes implies that a superpotential is generated for the Calabi-Yau complex structure moduli. In string theory this comes from a 3-point amplitude. In this section we derive this using the $G_3$ background discussed above. To this end, we briefly review how one defines string calculations in the hybrid formalism (see for example \cite{Berkovits:1996cr,Berkovits:2001nv,Linch:2006ig}) and then give the calculation for the superpotential.

One begins by twisting the superconformal algebra so that the operators in the 0-mode measure have weight 0. This twist is given by $(h_L,h_R)\rightarrow (h_L-\frac{q_L}{ 2},h_R-\frac{q_R}{ 2})$ and introduces a $+2$-charge anomaly in the left- and right-moving sectors, which is to be canceled by the measure. Next, we define the measure on the internal space. By observing that
\be
i\int_{CY_3} \Omega\w \bar\Omega = \frac{4}{3} \Vol (CY_3),
\ee
we define the internal part of the measure to satisfy
\be
\langle \Omega_L^+ \Omega_R^-\rangle_{CY_3} = 1.
\ee
For the 4-dimensional space-time, momentum conservation is implemented as usual by integrating over $d^4 x$. We also integrate over the zero modes of the $\th$ coordinates. Finally, we must incorporate the chiral bosons. The final result is
\be
\langle \th_L^2 \thab_L^2 \tha_R^2 \thab_R^2 e^{-\rho_L-\rho_R} \Omega_L^+ \Omega_R^-\rangle = 1.
\ee

In a Calabi-Yau background there are no chiral interactions or
superpotential for the massless moduli (see for example
\cite{Dixon:1987bg}). However, in the conformally CY background discussed
above, we find that there are scattering amplitudes which, in the low energy
field theory, are given by a superpotential. For example a scattering
amplitude\footnote{Normally such an amplitude would vanish by
$SL(2,{\mathbb C})$ and supersymmetry.
}
\be
\langle \cV^q \cV^r\rangle_{S_0+\delta S_{G_3}}=\langle \cV_{G_{(2,1)}} \cV^q \cV^r \rangle_{S_0} \ne 0
\ee
is now non-zero due to the flux background. Note that only vertex operators corresponding to complex
structure moduli feel the presence of this flux. As computed in \cite{Linch:2006ig} this amplitude gives rise to a superpotential
\be
\mathcal{W} = 3\int d^4x d^2\th^- g_s G^p_{(2,1)} \cV^q(x,\th^-) \cV^r(x,\th^-) C_{pqr},
\ee
where $C_{pqr}$ are $h^{2,1}$ intersection numbers, and we integrate over the unbroken space-time supersymmetry $d^2\th^-=d\th^- d\thab^-$. Here, $\cV^p$ are the massive-moduli chiral superfields reduced to $N=1$:
\begin{eqnarray}
\Vext &=& \phi^p + \th^- \chi^p
+(\theta^-)^2X^p_{--}+\dots,
\end{eqnarray}
where $\chi^p$ is the modulino corresponding to the complex structure
modulus $\phi^p$. Thus, we see that some of the complex structure moduli
get masses depending on which fluxes are turned on. This is as expected
from a supergravity analysis. At large volume, these amplitudes and
superpotential are packaged into the usual supergravity
superpotential\footnote{Here $\delta\Omega$ means the variation of
$\Omega$ induced by the presence of modulus fields considered when computing the
amplitude. It should be remembered that vertex operators only contribute
to {\it variations} of the complex and K\"ahler structures.}
\be
\mathcal{W} = \int G \w \delta\Omega.
\ee
We thus see that the hybrid easily calculates the quadratic contribution of the GVW superpotential at large volume.

\section{Conclusion and Future Directions}
\label{sect:conclusions}
In this paper we have studied type IIB string compactifications to 4 dimensions. We identified hybrid formalism vertex operators corresponding to NSNS and RR fluxes on the internal space. We showed that in flux perturbation theory one may easily compute various known physical quantities such as warping and the generation of a superpotential for the massive moduli. The manifest space-time supersymmetry automatically gives the supersymmetry conditions on the flux.

There are many interesting future directions to pursue. For example, in forthcoming work\cite{Linch:2008a} we show that in the hybrid model the presence of RR flux implies that the space-time supersymmetry algebra develops a central charge. In the same work, we also show that RR flux naturally leads to non-anti-commutativity in space-time. This has been discussed previously\cite{Seiberg:2003yz,deBoer:2003dn}, and we find a natural analogue here. Finally, with an eye toward proving the existence of flux backgrounds, we have developed a formulation of the hybrid worldsheet theory in terms of RR ground states. Following \cite{Rohm:1985jv}, it can be argued that in the Born-Oppenheimer approximation flux vacua are stable against perturbative $\alpha'$ corrections in RG flow, giving evidence for the existence of flux vacua as string solutions.

There are other directions to pursue. In the context of flux compactifications one can study the analogue of the gauged linear sigma model, first developed in \cite{Witten:1993yc}, with RR flux deformations\cite{McOrist:2008b}. One can also easily study D3-brane backgrounds in this formulation. One very interesting direction is to understand orientifolding, and thereby open the realm of compact solutions to analysis. Of course, the ultimate goal of this line of work is to develop methods of calculation to the point that one is able to study string solutions that are string scale. This would give insight into the properties of string vacua not attainable in supergravity and an understanding of the landscape of perturbative type II string vacua.

\acknowledgments
We would like to thank Nick Halmagyi, Oleg Lunin, Savdeep Sethi,
Alessandro Tomasiello and especially Ilarion Melnikov and Warren Siegel for helpful and
insightful comments and discussions. We would also like to thank Ilarion Melnikov for a careful reading and comments on the manuscript. We also acknowledge the organizers of the Simons 2007 workshop in which this work was initiated.
The work of WDL3 is supported by NSF grant numbers PHY 0653342 and DMS
0502267. JM is supported by a Ledley Fellowship. The work of BCV is
supported by FAPESP grant number 06/60175-5.
\appendix

\baselineskip=18pt


\providecommand{\href}[2]{#2}\begingroup\raggedright\endgroup
\end{document}